\documentclass[twocolumn,amsmath,amssymb,pra]{revtex4}
\makeatletter
\usepackage{graphicx}
\usepackage{epsfig}
\usepackage{epsf}
\usepackage{amssymb,amsmath}
\usepackage{amsmath}
\usepackage{ifthen}
\usepackage[bookmarks=true]{hyperref}
\newcommand{\half}{\frac{1}{2}}
\newcommand{\ket}[1]{\vert{#1}\rangle}

\newcommand{\Tr}[1]{\textrm{Tr}\left[{#1}\right]}
\newcommand{\atan}[1]{\textrm{atan}{#1}}
\newcommand{\ham}{\mathcal{H}}

\begin{document}

\title {Quantum correlation in disordered spin systems:\\
entanglement and applications to magnetic sensing}
\author{P. Cappellaro and M.~D. Lukin}
\affiliation{ITAMP -- Harvard-Smithsonian Center for Astrophysics and  Physics Department, Harvard University, Cambridge, MA 02138, USA}

\begin{abstract}
We propose a strategy to generate a many-body entangled state in a collection of randomly placed, dipolarly coupled electronic spins in the solid state. By using coherent control to restrict the evolution  into a suitable collective subspace, this method enables the preparation of GHZ-like and spin-squeezed states even for randomly positioned spins, while in addition protecting the entangled states against decoherence. We consider the application of this squeezing method to improve the sensitivity of nanoscale magnetometer based on  Nitrogen-Vacancy spin qubits in diamond.
\end{abstract}

\maketitle

\section{Introduction}
Entangled states have attracted much interest as intriguing manifestation of non-classical phenomena in quantum systems. The creation of a many-body entangled state is a critical requirement in many quantum information tasks, such as quantum computation and communication, as well as in measurement devices. Here we  outline a novel approach to obtain  many-body entangled states in a solid-state system of dipolarly coupled electronic spins. In order to achieve the generation of entanglement in the presence of disordered couplings we take advantage of the fine experimental control reached by magnetic resonance to constrain the evolution to a suitable collective subspace~\cite{Rey08}. Furthermore,  this restriction to a collective subspace protects the entangled state from decoherence, thus bringing into experimental reach a particular class of entangled states (the spin squeezed states) that are of great practical interest. Spin squeezing in  solid-state systems could have an immediate application to improve the sensitivity of recently demonstrated spin-based magnetometers~\cite{Taylor08,Maze08,Balasubramanian08}. We show that controlling the naturally-occurring interactions to obtain a desired entangled state could yield a high sensitivity magnetometer in a nano-sized system for high-spatial resolution.

The paper is organized as follows. We first describe in section II entanglement generation in ideal and disordered systems, outlining the control techniques required to achieve the projection of the evolution to the desired subspace and its regimes of validity  for different geometry distributions of the spins. In section III we then apply the method to spin squeezing and we show in section IV how the projection is also capable of reducing the noise effects, thus making squeezing advantageous for metrology.  Finally in section V we present a possible implementation of the squeezing scheme. We focus our analysis on a system based on spin defects in diamond (Nitrogen-Vacancy center, NV~\cite{Childress06Et,Dutt06Et,Gaebel06}, Fig. 1 a). The NV electronic spins can be optically polarized and detected, and exhibit excellent coherence properties even at room temperature, allowing for a remarkable combination of sensitivity to external magnetic fields and high spatial resolution. We  describe the operating regime  of a spin squeezed NV magnetometer and the achievable sensitivity improvement. 
We emphasize that the described techniques are applicable to other spin systems, such as other paramagnetic impurities or trapped ions~\cite{Rey08}.

\section{Entanglement generation}
\begin{figure}[hb]
	\centering
		\includegraphics[scale=0.5]{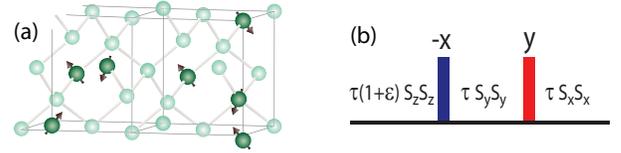}
	\caption{(a) System model: crystal with randomly placed electronic spins. (b) Control sequence: The Ising interaction is rotated along  three axis to yield an isotropic interaction; adjusting the time delays, a small perturbation along the z direction is retained, to obtain the 1-axis squeezing operator.}
	\label{fig:Model}
\end{figure}
\subsection{Entanglement in ideal and disordered systems}
We  consider a solid-state system of N spin particles
with two relevant internal states (0,1), each described by Pauli matrix operators $\sigma_\alpha^k$.  Interactions among the spins can be used to generate entanglement. In particular, evolution of an initially uncorrelated state under the so-called one-axis squeezing Hamiltonian $\ham^{1a}_{sqz}=dJ_z^2$ is known to create the multi-spin GHZ state (here we introduce the collective operator $J_\alpha=\sum_kS_\alpha^k$, with $S_\alpha^k=\half\sigma_\alpha^k$). Starting from the fully polarized state along the $x$-direction $\ket{N/2,N/2}_x=\sum_{m_z,\mu}C_{m_z,\mu}\ket{N/2,m_z,\mu}_z$, the different $m_z$-components acquire  $m_z^2$ dependent phases that lead to collapse and revivals of the collective polarization $J_x$. At a time $t=\pi/(2d)$ the system is found in the collective GHZ state, 
 $\ket{\psi_{GHZ}}_x=\frac1{\sqrt{2}}(\ket{N/2,N/2}_x+(-i)^{N+1}\ket{N/2,-N/2}_x)$. 

\begin{figure*}[t]
	\centering
		\includegraphics[scale=0.5]{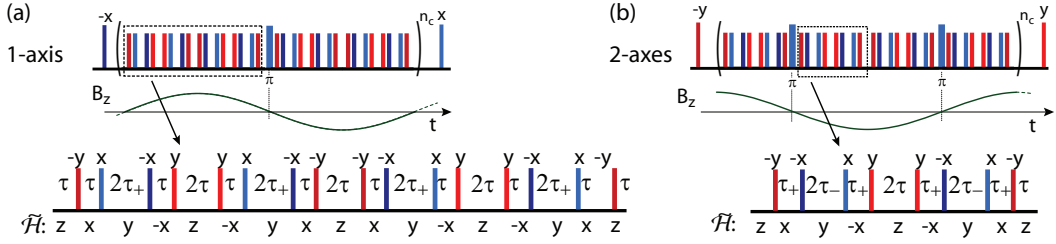}
\caption{ (Color online) Decoupling sequence. The narrow bars are $\pi/2$ pulses around different axis in the $m_s=\pm 1$ manifold, while the wide bars are $\pi$-pulses. The overall pulse sequence  comprises  4 MREV8 sequences embedded in a spin echo sequence, with 34 pulses and a cycle time of $48\tau$. 
By varying the  length of the time delays and the pulse phases, we obtain the squeezing Hamiltonians. The modified intervals are indicated by $\tau_+$ and $\tau_-$. For the one-axis squeezing (a) we obtain the first order Hamiltonian  $\overline{\ham}=(\ham_{H}+\epsilon\ham_{zz})/3$ (neglecting terms $\propto \epsilon\sin{\nu}^2$) and the linear Hamiltonian $S_z\rightarrow \sqrt{2}(S_z\cos{\nu}+S_y\sin{\nu})/3$. For the 2-axes case (b) the first order Hamiltonian is $\overline{\ham}=(\ham_{H}+\epsilon\ham_{DQ})/3$ and the effective field is $(S_y-S_x)/3$. $\tilde{\ham}$ is the \textit{direction} of the internal Hamiltonian in the interaction frame. 
$B_z$ is the AC field to be measured.}
	\label{fig:MREV8}
\end{figure*}
In most physical systems, however, the interactions among spins are not of the type described by the ideal entangling operator. Quite generally the Hamiltonian can be written as $\ham_{zz}=\sum_{lj}d_{lj}S_{z,l}S_{z,j}$, where the couplings $d_{lj}$ are effectively limited to a finite number of neighbors. If it were possible to precisely engineer the strength of the couplings or the graph connectivity of the spins, it would still be possible to obtain a maximally entangled state such as the GHZ state~\cite{Christandl05,DiFranco08a}. For example, even in the limit of nearest neighbor couplings only, a particular choice of couplings ($d_{k,k+1}=2d_0\sqrt{k(N-k)}/N$, with $d_0$ the maximum coupling strength) in the presence of a spatially-varying magnetic field $\ham_X=\sum_k \sqrt{(2k-1)(2N-2k+1)}S^k_x$ creates the $N$-spin GHZ state in a time $d_0t=N\pi/8$.

In naturally occurring spin systems, where  the couplings are usually given by the dipolar interaction scaling as $1/r^3$ with distance, it is difficult to engineer the couplings in the desired way and one has to deal with a disordered set of coupling strengths. 
We assume here an Ising interaction, $\ham_{zz}$, with $d_{lj}=\frac{\mu_0g^2\mu_D^2}{4\pi\hbar}\frac{(3\hat{r}_{lj}.\hat{z}_l\hat{r}_{lj}.\hat{z}_j-1)}{r_{lj}^3}$. This is the case for an ensemble of NV electronic spins, where the zero-field splitting Hamiltonian $\Delta S_z^2$ is the largest quantizing energy scale (here we assume to operate in the $\pm1$ manifold only, that constitutes an effective spin-$\half$ system).
This Hamiltonian  will still generate entangled states, but the amount and type of entanglement may not be as desired. 

\subsection{Creating a global Hamiltonian}
To obtain the desired high entangled state, we propose to create an effective collective Hamiltonian starting from a local one~\cite{Rey08}. 
Notice that the projection of the $\ham_{zz}$ Hamiltonian onto the $J=N/2$ subspace is given by:
\begin{equation}
	P_{N/2}[\ham_{zz}]=\frac{ D}{(N-1)}(J_z^2-N\openone),
\end{equation}
with $D=\frac{1}{N}\sum_{l,j}d_{lj}$. This operator can create the GHZ state, at the expenses of an increased evolution time. The restriction to the maximum angular momentum manifold can be achieved if $\ham_{zz}$ is only a small perturbation to a stronger Hamiltonian that conserves the total angular momentum. In the following, we will show that with collective coherent control techniques it is possible to let the system evolve under the  interaction:
\begin{equation}	\ham_H^{1a}=\epsilon\ham_{zz}+\ham_{H}=\epsilon\sum_{lj}d_{lj}S_{z,l}S_{z,j}+\sum_{lj}d_{lj}\vec{S}_{l}\cdot\vec{S}_{j}
\label{Ham}
\end{equation}
If $\epsilon\ll1$, the Ising  Hamiltonian is just a perturbation to the isotropic Heisenberg Hamiltonian $\ham_H$, and to first order approximation, we only retain its projection on the $J=N/2$ manifold. 
Notice however that the squeezing Hamiltonian strength is now  $\frac{\epsilon~D}{N-1}$, so that it is necessary to apply the squeezing interaction for a time increasing with the number of spins: $t\approx \frac{\pi}2\frac{N}{\epsilon D}$. 

To obtain the desired Hamiltonian $\ham_H^{1a}$, we propose to apply control techniques based on fast modulation of the internal Hamiltonian by cyclic sequences of pulses. These techniques  have been  used in NMR to obtain a wide range of desired interactions (see Appendix A).  
By cyclically rotating the Ising interaction among three perpendicular axis, it becomes  on average isotropic (Fig. 1-b) . More complex modulation sequences (such as {\sc Mrev}8~\cite{Rhim73,Mansfield71}), achieve the averaging to higher order in $\left\|\ham~t\right\|$. 
By a careful adjustment of the delays between pulses, it is possible to retain part of the $\ham_{zz}$ Hamiltonian, so that the effective Hamiltonian is $\overline{\ham}=\frac13\ham_H^{1a}$ (see Fig. \ref{fig:MREV8}.a). 

The validity of the approximation taken in considering only the projection of $\ham_{zz}$ onto the ground state $J=N/2$ manifold relies on the existence of an energy gap between the ground state manifold and the $J<N/2$ manifolds, induced by the isotropic interaction $\ham_H$. The magnitude (and existence) of this gap depends on the geometry and spin-spin couplings of the system. For a 1D system with  constant nearest-neighbor couplings $d_0$ the energy gap $E_g$ decreases as $d_0/N^2$, thus we must take $\epsilon\propto1/N^2$ to always remain in the regime of validity of the approximation: the time required to achieve the GHZ state increases rapidly with the number of spins. The nearest-neighbor, 1D model is the worst case scenario; more generally, the time required will be a function of the dependence of $E_g$ and $D$ on $N$. For example, for a dipolarly coupled regular 1D system, $D=\zeta(3)d_0$ and the gap scales as $E_g\sim (d_0/N^2)\log{N}$, with $d_0$ the coupling at the minimum distance $r_0$. 

Better scaling can be achieved in a quasi-2D system, consisting of layers of spin impurities. Already a nearest-neighbor square lattice will have  $E_g\sim d_0/N$ (it is in general $E_g\sim N^{-2/dim}$, where $dim$ is the system dimensionality), while for couplings decaying with distance as $r^{-3}$, $D\approx 7d_0$ and the gap is $E_g\sim d_0/\sqrt{N}$. In this case, the evolution time must increase with the spin number as $t\sim N^{3/2}$. A similar scaling is predicted if the spins have  random spatial locations with density $n_s$. The gap can be estimated from the minimum coupling strength $E_g\sim d_{min}\frac N2$ with $d_{min}\sim (\frac{n_s}{N})^{3/dim}$ (where we assumed all the couplings to be positive). Thus we obtain again the same scaling of the gap energy $E_g\sim1/\sqrt{N}$ in 2D and a constant gap in 3D. This last result, however, must be taken with caution, since the angular dependence of the dipolar couplings in 3D unavoidably gives rise to negative couplings; in that case not only the gap could even disappear, but the average coupling strength $D$ is zero for an isotropic spatial distribution. For finite size systems, however, because of the large variance of the couplings, a better estimate for $D$ is given not by the average but by the median of the dipolar coupling: $D\approx \textrm{median}\left[(3\cos{\vartheta}^2-1)d_0r_0^3/r^3\right]\approx\frac{2\pi}{3}d_0r_0^3\frac {n_s}{N+2}$, and we expect to obtain a non-zero gap with high probability. Still, the times required to obtain the GHZ state increase rapidly with the spin number in all the possible configurations presented. We thus turn our attention to a different class of entangled states, whose preparation time under ideal conditions decreases with the number of spins. A set of states that possess this property are the so-called spin squeezed states, which are of particular interest in metrology tasks.

\section{Spin squeezing}
Spin squeezed states are many-body states showing pairwise entanglement~\cite{SandersSqzEnt} and reduced uncertainty in the collective spin moment in one direction~\cite{Kitagawa}.
This reduction in the measurement uncertainty,  achieved without violating the minimum uncertainty principle by a redistribution of the quantum fluctuations between non-commuting variables, can be exploited to perform metrology beyond the Heisenberg limit.  
 One-axis twisting ($\ham^{1a}_{sqz}$) and two-axis twisting [$\tilde{\ham}^{2a}_{sqz}=i~d(J_+^2-J_-^2)/2$] Hamiltonians have been proposed to achieve this goal~\cite{Kitagawa}.

The degree of squeezing of a spin ensemble is evaluated by the squeezing parameter $\xi$. Several definitions have been proposed, depending on the context~\cite{PRA033611}. If the focus is simply to describe a non-uniform distribution of the quantum fluctuations, the appropriate quantity is $\xi_h=\Delta J_i/\sqrt{J_j/2}$, where $\Delta J_\alpha$ is the uncertainty in the $\alpha \in \{x,y,z\}$ direction of the collective angular momentum and $J_\beta$ its expectation value in a different direction. When spin squeezing is instead used in the context of quantum limited metrology~\cite{Wineland94}, the squeezing parameter should measure the improvement in signal-to-noise ratio  for the measured quantity $\varphi$, 
\begin{equation}
	\xi=\Delta\varphi_{sqz}/\Delta\varphi_0=\sqrt{N }\frac{\Delta J_z}{\|\frac{\partial \langle J_z\rangle}{\partial \varphi}\|}
\end{equation}
This definition is associated to Ramsey-type experiments, in which an external magnetic field is measured via the detection of the  accumulated phase $\varphi$ due to the Zeeman interaction and the phase uncertainty for a product state is $\sim1/\sqrt{N}$.  

In the limit of large spin numbers, using the one and two-axis squeezing operator, the optimal squeezing parameters are $\xi_{1a}=\frac{3^{1/3}}{\sqrt{2}N^{1/3}}$, at a time $t_{1a}=\frac{3^{1/6}}{d~N^{2/3}}$~\cite{Kitagawa}, and $\xi_{2a}\sim\sqrt{\frac{1+2\sqrt{3}}{2N}}$ at $t_{2a}\approx\frac1{d~N}\log{\frac{2N}{\sqrt{3}}}$~\cite{AndreLukin,GeremiaStockton}, respectively.
The one-axis squeezing operator reduces the variance of the collective magnetic moment along a direction at a variable angle $\nu\approx0$ in the y-z plane, while for the two-axis operator the uncertainty reduction is in the x-direction.

An arbitrary Hamiltonian $\ham_{zz}$ can generate a squeezed state~\cite{Wang01}, although the squeezing would be less than in the ideal case and it becomes difficult to predict the optimal squeezing time and direction. For example, in the limit where the interaction is limited to first neighbors, the maximum squeezing achievable is fixed (independent of N) and bounded by $\xi_{nn}\approx 0.73$~\cite{Sorensen01}. 
We can nonetheless apply the techniques presented in the previous section to project out the ideal squeezing Hamiltonian from the natural occurring disordered interaction. 

The same control techniques described above can also generate two-axis squeezing. A different choice of time delays (Fig. \ref{fig:MREV8}.b) will produce the Hamiltonian:
${\ham}_H^{2a}=(\epsilon\ham_{dq}+\ham_H)/3$, 
where we introduced the so-called double quantum Hamiltonian  $\ham_{dq}=\sum_{lj}d_{lj}(S_{x,l}S_{x,j}-S_{y,l}S_{y,j})$.
This operator creates squeezing by a two-axis twisting mechanism: for $\epsilon\ll1$, we only retain the projection of $\ham_{dq}$ on the $J=N/2$ manifold, $P[H_{dq}]=\frac{ D}{(N-1)}[(J_x^2-J_y^2)-N\openone]$, which is equivalent to $\ham_{sqz}^{2a}=d(J_x^2-J_y^2)/2$ (a 2-axis squeezing operator with optimum squeezing direction  along the $\pi/4$ axis in the x-y plane). With this operator a squeezing parameter $\xi_{2a}\sim 2/\sqrt{N}$ can be achieved in a time $t_{2a}\approx\frac{(N-1)}{\epsilon D~N}\log{\left(\frac{2N}{\sqrt{3}}\right)}$: notice that the two-axis squeezing operator provides not only a better optimal squeezing, but also in a shorter time than the one-axis operator ($t_{1a}\approx \frac{3^{1/6}}{\epsilon D}N^{1/3}$). 
It is also possible to embed the control sequence within a spin-echo scheme, as needed for the control of dephasing due to a quasi-static spin bath~\cite{Taylor08}, as well as to adjust the pulse phases in order to obtain an average external field operator acting on the direction of maximal squeezing. 

\section{Protection against the noise}
Entangled states are known to decay more rapidly due to dephasing than separable  states, so that in practice the improvement in sensitivity is often counterbalanced by the need to reduce the interrogation time~\cite{Huelga,auzinsh04}. In particular, if the system is prepared in the maximally entangled state (an N-particle GHZ state) the sensitivity improvement is completely lost in the presence of some classes of decoherence (single particle dephasing)~\cite{Huelga}, as the system is $N$-time more sensitive both to the signal and the noise. 
A partially entangled state, such as a spin squeezed state, can instead provide an advantage over separable states~\cite{KitagawaSqueeze}. 

  In the solid-state systems here considered the source of decoherence is usually the interaction of the electronic spins with other paramagnetic impurities and with nuclear spins. In the case of the NV electronic spins, the noise is caused by nitrogen electronic spins and by $^{13}$C nuclear spins. Since the couplings to these spins are much smaller than the zero-field splitting, they can only cause dephasing but not spin flips. 
  The noise can be modeled as a fluctuating local magnetic field and represented by a single-spin dephasing stochastic Hamiltonian, $\ham_{noise}=\sum_k\omega_N^k(t)S_z^k$; $\omega_N^k(t)$ are assumed to be independent stationary Gaussian random variables with zero mean and correlation function $\overline{\omega_N^k(t_1)\omega_N^h(t_2)}=\Omega_k f_k(t_2-t_1)~\delta_{k,h}$. This noise leads to a decay of the average magnetization in the transverse direction: $\overline{\langle J_x(t)\rangle}=e^{-N\Gamma t}J_x^{id}(t)$ with respect to the ideal case $J_x^{id}(t)$. Here we  defined $\Gamma=\frac 1{2t}\int_0^t dt_1 \int_0^t dt_2 \overline{\omega_N(t_1)\omega_N(t_2)}$ and the average noise: $\omega_N(t)=1/N\sum_k\omega_N^k(t)$. 
  This decay affects both   product and 1-axis squeezed states in the same way (since the squeezing operator commute with the noise). The angular momentum uncertainty for the squeezed state is also affected, so that the squeezing parameter has now the value:
\begin{equation*}
	\xi^2_{1a}=\frac{1+\frac{N-1}4e^{-N\Gamma t}[(1-e^{-N\Gamma t})+1/R]P(R-1)}{\cos{(dt)}^{2N-2}}
\end{equation*}
where $R=P/\sqrt{P^2+Q^2}$ [with $P=1-\cos^{N-2}{(2dt)}$ and $Q=4\sin{(dt)}\cos^{N-2}{(dt)}$, see appendix B].

\begin{figure}
	\centering
		\includegraphics[scale=0.32]{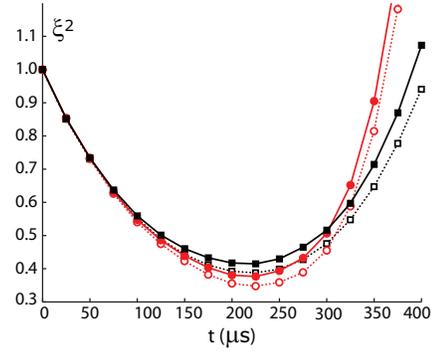}
	\caption{(Color online) Simulations (8 spins) of the squeezing parameter squared $\xi^2$ under 1-axis (squares) and 2-axis (circles) squeezing control schemes, with (solid lines) and without (dotted lines) the presence of noise.  
The noise was modeled by a random magnetic field in the z direction acting independently on each spin; noise parameters where $\Gamma=3$kHz, $\tau_c=100\mu s$. The spins were distributed randomly on the lattice of a quasi-2D diamond slab of depth 9nm and area $30\times30$ nm. The density was $ 10^{18} \textrm{cm}^{-3}$  resulting in $D=.4$ MHz and $E_{gap}=50$kHz. The control sequence of Fig. \ref{fig:MREV8} was used, with  time delays of 1.4$\mu s$ and 1 to 12 cycles.}
	\label{fig:Simulation}
\end{figure}

For small enough noise, such that the optimal squeezing time is shorter than the decoherence time, the squeezing parameter scales with the number of spins and decay rate as  $\xi_{1a}\sim\frac{3^{1/3}}{N^{1/3}}\sqrt{\frac{1+(\Gamma/ \epsilon D)^2}2}$. Only if the decoherence rate is such that $\Gamma/\epsilon D\sim 1$ we retain the ideal  dependence of the squeezing parameter $\xi_{1a}\sim N^{-1/3}$. For stronger noise, the optimal time must be reduced and the maximum squeezing does not reach its optimal value.

The isotropic Hamiltonian $\ham_H$ offers protection against the noise, provided this is smaller than the gap to the $J<N/2$ manifolds~\cite{Rey08}.
To first order approximation, the only part of the noise operator that has an effect on the system is its projection on the $J=N/2$ subspace: $\ham_{noise}\rightarrow\omega_N(t)~J_z$. 
Not only the signal decays $N$-times slower, $\overline{\langle J_x(t)\rangle}= e^{-\Gamma t}J_x^{id}(t)$, but also the uncertainty is less affected. To provide a  fairer comparison, we calculate a squeezing factor in which the non-squeezed state is also protected by the isotropic Hamiltonian: 
\begin{equation*}
\begin{array}{ll}
	\xi^2_{1a}=&\cos{(dt)}^{2-2N}\left\{1+\frac{N-1}{4}e^{-\Gamma t}\right.\\&\left.\times[(1-e^{-\Gamma t})-2\sinh{(\Gamma t)}/P+1/R]P(R-1)\right\}
\end{array}
\end{equation*}

The optimal squeezing now scales as $\xi_{1a}\sim\frac{3^{1/3}}{\sqrt{2}N^{1/3}}+\sqrt{\frac\Gamma{N~d}}$. We obtain a lower bound for the decoherence rate, $\Gamma/\epsilon D=o[N^{1/3}]$, that still allows optimal squeezing. 

To take into account corrections to the truncation of the noise operator, we can follow the  analysis  in~\cite{Rey08} to find the leakage rate to the $N/2-1$ subspace. 
Because of the energy gap from the $J=N/2$ to the $J=N/2-1$ manifold (which can be populated by the flipping of one spin caused by the single-spin noise) the leakage outside the protected space is negligible, unless the energy gap to the first excitation is comparable with the cut-off energy $\omega_c$ of the noise (notice that $\omega_c=1/\tau_c$,  the noise correlation time.)	 

While there is no  analytical solution for the noise effects on the squeezing under the 2-axis Hamiltonian, we expect a similar advantage from the reduction of the noise to its collective part only. It is known that even the maximally entangled (GHZ) state does not exhibit a faster dephasing under collective noise~\cite{ChildsSqz}. Simulations for 8 spins show the expected improvement (Fig.  [\ref{fig:Simulation}]).

\section{Application to metrology: magnetic sensing and decoherence}
We now describe an application of electronic spin squeezed states to precision magnetometry. 
Electronic spins in the solid state can be used to sense external magnetic fields, by monitoring the Zeeman phase shift between two sublevels via a Ramsey experiment. For small phase angle (weak fields) the signal measured (the total magnetization in the field direction) is proportional to the field.

The ideal sensitivity to the measured magnetic field for $M=T/t$ measurements and $N$ spins is given by 
\begin{equation}
\Delta B_z=\frac{\hbar}{g \mu_B\sqrt{M}}\frac{\Delta J_z}{\|\frac{\partial \langle J_z\rangle}{\partial B_z}\|}=\frac{\hbar}{g \mu_B} \frac{1}{\sqrt{N~t~T}}\xi(t,N)	
\end{equation}

For sake of concreteness, we consider the recently proposed diamond-based magnetometer~\cite{Taylor08,Maze08,Balasubramanian08} (Fig. 4-a). The electronic spin associated to the NV in diamond is a sensitive probe of external, time-varying magnetic fields, due to its long coherence time and  optical detection~\cite{Childress06Et,Dutt06Et}. 
The electronic spin triplet can be polarized  under application of green light and controlled by ESR pulses even at zero external magnetic field, thanks to the large zero-field splitting ($\Delta=2.87$GHz) (Fig. 4-b). In order to increase the interrogation time, a spin-echo based operating regime has been proposed, thus making the magnetometer sensitive to AC fields. The operating scheme of the magnetometer is depicted in Fig. 4. High spatial resolution is achieved by using for example a crystal of nanometer scales, which could be operated as a scanning tip. Sensitivity can be improved by increasing the number of NV's in the tip, but this comes at the cost of errors introduced by the NV-NV couplings. Instead of refocusing these couplings~\cite{Taylor08}, here we propose to use them to create a squeezed state. 
\begin{figure}[tb]
	\centering
		\includegraphics[scale=0.55]{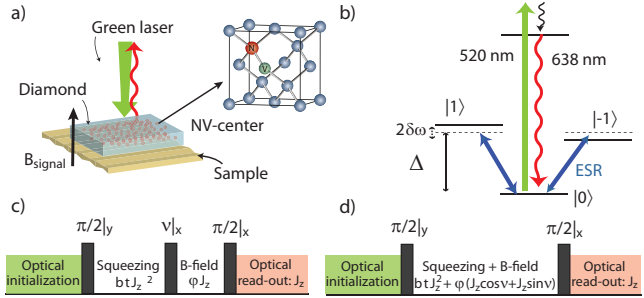}
	\label{fig:NVScheme}
	\caption{Magnetometer measurement scheme. a) A possible magnetometer setup, with NV center implanted in a slab of diamond. b) Optical and microwave control of a single NV center. Squeezing control sequences: c) A squeezed state is prepared before applying an external magnetic field to measure. d) Squeezing and Ramsey experiment are done simultaneously}
\end{figure}

\begin{figure*}[htb]
	\centering
		\includegraphics[scale=0.3]{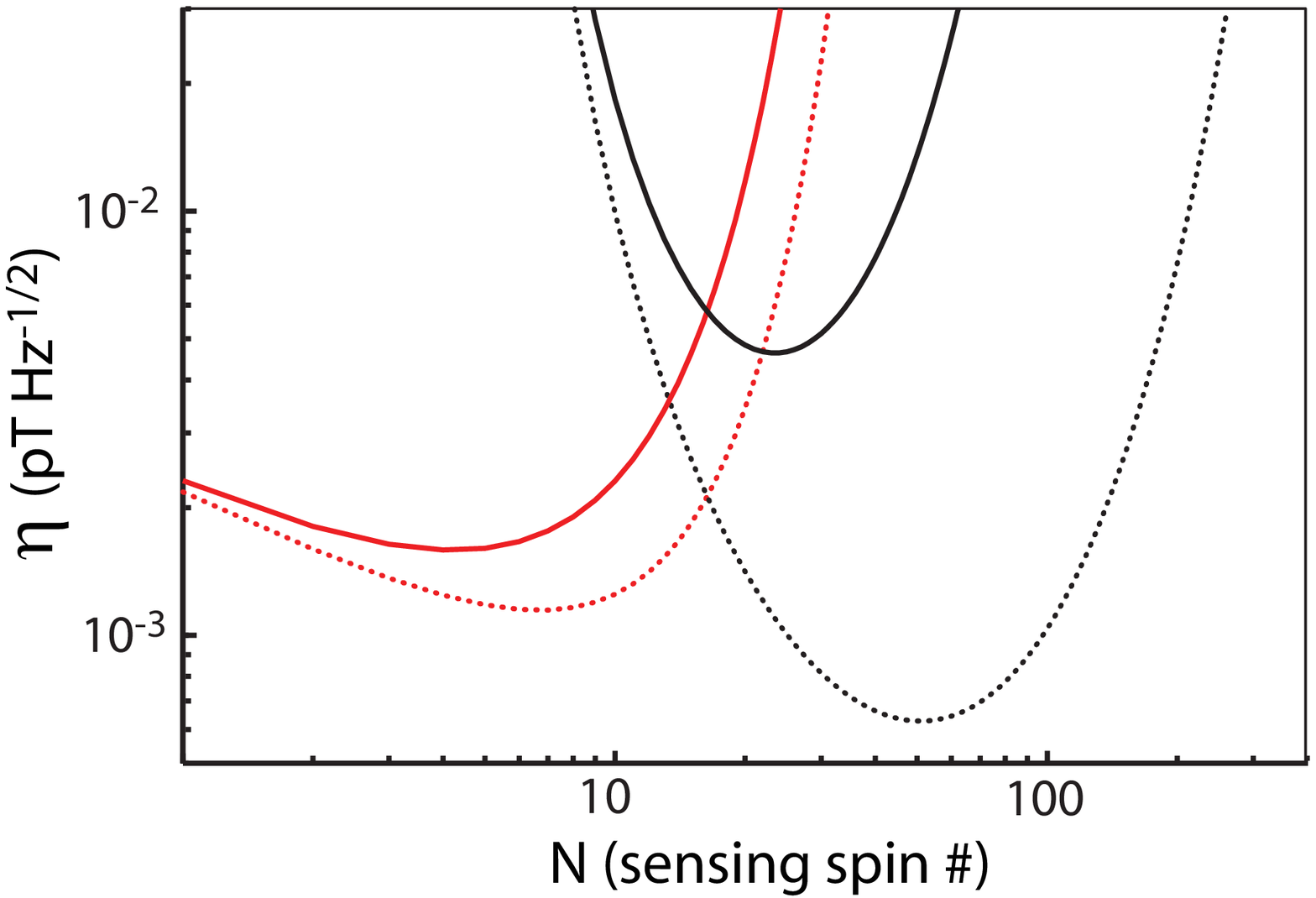}
		\ \ \ \ \ 
		\includegraphics[scale=0.3]{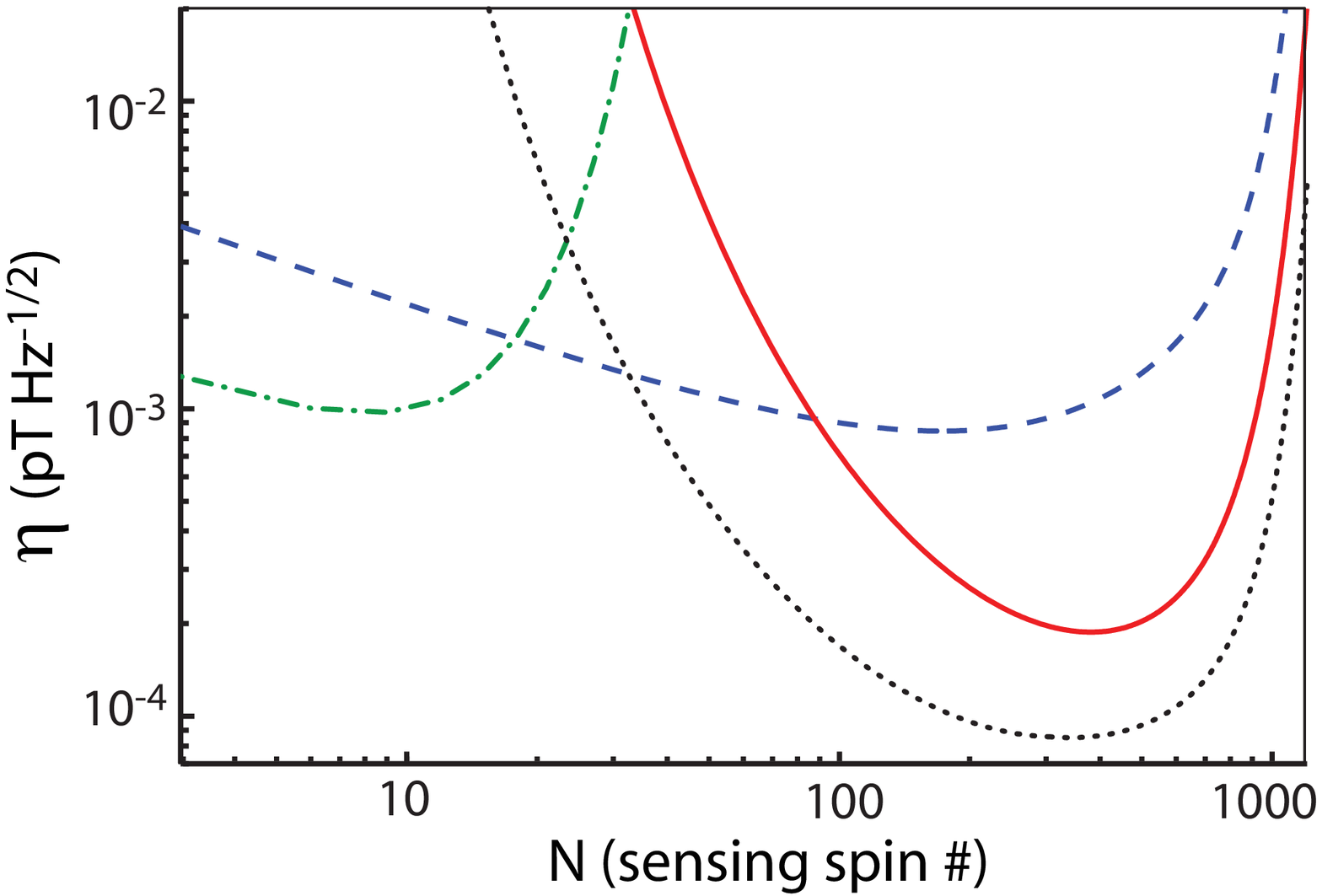}
	\caption{Sensitivity to an AC field with 22kHZ frequency, in a quasi 2D geometry [$V=(300 \textrm{nm})^2\times 10\textrm{nm}$].
	Left: Sensitivity for conversion efficiency $f=50\%$ (dotted lines) and $f=23\%$ (solid lines). Notice that only for $f\gtrsim 50\%$, the 2-axis squeezing approach (black lines) is preferable to the echo-only sequence (red lines).
	Right: Sensitivity for $f=.90$ (or equivalently, for a 30-fold increase in $T_2$ time thanks to refocusing, with respect to the $T_2$ at the highest conversion achieved~\cite{Wee07,Acosta09x}). 
	Green line (dash-dotted): Sensitivity under spin-echo, without any refocusing of the spin-spin coupling.
	Blue dashed line: Sensitivity under {\sc Mrev}8 sequence.
	The control sequence was repeated $n_c=6$ times with a delay between pulses of $\delta\tau=1.5\mu$s.
	Red line: Sensitivity under 1-axis squeezing, obtained with the same sequence parameters as above. 	Black dotted line:  Sensitivity under 2-axis squeezing.
	The time delays $\tau_+,\ \tau_-$ (see Fig. 2) were chosen in order to get an $\epsilon$ as small as needed to restrict the evolution to the protected manifold and the number of cycles increased as needed to obtain the optimal squeezing time.
	}
	\label{fig:etaSqz}
\end{figure*}
Unfortunately, the coherence properties of high NV-density diamonds or nanocrystals~\cite{Maze08} are currently worse than for bulk, pure diamonds.
NV's are usually created by nitrogen implantation, followed by annealing that makes vacancies migrate and combine to the nitrogens~\cite{Meijer05,Rabeau06}. The current conversion efficiency $f$ is quite low (between 10\% and 23\%~\cite{Wee07}), thus interactions with the epr bath (in particular P1 nitrogen centers~\cite{Hanson08}) limit the coherence time and bound the allowed NV densities. The spin-echo signal is a function of  the root-mean-square coupling among impurities and it decays exponentially as $e^{-t/T_{epr}}$.  We find $T_{epr} \approx 4/[\frac{\mu_0}{4\pi}\frac{(g\mu_B)^2}{\hbar} n_{epr}]$~\cite{Taylor08}, where the density of paramagnetic impurities is $n_{epr} = n_s(1-f)/f$ (with $n_s$ the  NV center density). Not only the coherence time is short, but also the internal dynamics of the epr bath is fast, so that the protection provided by the gap created with the introduction of the isotropic Hamiltonian is not effective, as the condition $\omega_c\ll E_{gap}$ is no longer valid.
Improved implantation schemes~\cite{Prawer06}, coherent control techniques~\cite{Taylor08}, polarization of the nitrogen,  either at low temperature~\cite{Takahashi08} or by optical pumping, can reduce the effects of epr spins, up to the point where NV-NV couplings are the most important interactions. 
This is the regime suitable for squeezing. A conversion efficiency of 50\% (or equivalently a 3 fold increase in the relaxation time due to the paramagnetic impurities) is needed to start seeing an advantage of the squeezing scheme over a simpler scheme based on repeated echoes (CPMG sequence~\cite{Meiboom58}) as shown in Fig.~\ref{fig:etaSqz}(a). 

\section{Implementation in a diamond nano-crystal} 
If the material properties of NV-rich diamonds can be improved, a sizable squeezed state could be obtained in a nanocrystal (or in a suitably implanted portion of a bulk diamond, if surface effects are to be avoided). 
The residual decoherence mechanism is due to couplings  to the nuclear spin bath (1\% abundance of $^{13}$C), while  spin relaxation occurs on timescales much longer than milliseconds and is thus neglected. The resultant decay of the signal, with a $T_2$ time on the order of $200-600\mu$s~\cite{Takahashi08,Maze08}, is about the same for a product state and for a squeezed state protected by the isotropic interaction.  
Errors in the creation of the average squeezing Hamiltonian $\bar{\ham}$ by the multiple pulse sequence can be taken into account as a decaying term calculated from the third order average Hamiltonian~\cite{Haeberlen76} and a  moment expansion to second order~\cite{Mehring} (see Appendix A). For the {\sc Mrev}8 sequence of Fig.~\ref{fig:MREV8}, we obtain a decay term $\exp{\left(-{\tilde{\alpha}n_s^6 t^2}\right)}$, where $t$ is the total averaging time and the coefficient $\tilde{\alpha}\sim(\frac{\mu_0}{4\pi}\frac{(g\mu_B)^2}{\hbar})^6 \tau^4$ depends on the actual NV-NV couplings ($\tau$ is the delay between pulses, see Fig.~\ref{fig:MREV8}). 

Taking into account the described effects, as well as the scaling of the field due to the control sequence and  subunit efficiency $C$ of the optical read-out process, the sensitivity per root averaging time $\eta=\Delta B_z\sqrt{T}$ is:
\begin{equation}
	\label{sensitivityMREV8}
	\eta= \frac{\hbar}{g \mu_B} \frac{3 \pi e^{[T/T_2]^3} e^{T/T_{epr}}
       }{C \sqrt{2n V~t}}e^{\tilde{\alpha}n_s^6 T^6}\xi(t_{sqz},n_s V)
\end{equation}
where $T=t+t_{sqz}$ is the total experimental time and $t$ is chosen to maximize the signal $t=t_{opt}$.

In order to reduce the total experiment time, to avoid decoherence, we would like to create squeezing while applying the external magnetic field. The total time would then be $T={max}(t_{sqz},t_{opt})$ instead of their sum and the interrogation time  $t=T$ (See Fig. 4 c-d).
For the 1-axis squeezing, if $\nu_{opt}$ were $0$, we could squeeze the spins while acquiring a phase from the external field since the two Hamiltonians would commute. More generally, since $Dt_{1a}$ and the accumulated phase $\varphi$ due to the external field are both small, we can approximate the desired evolution as
\begin{equation*}
\begin{array}{l}
	e^{-i\varphi J_z}e^{-i\nu J_x}e^{-i\ham_{sqz}^{1a}t_{1a}}\approx \\
	e^{-i\nu J_x} 	
\exp{\left[-i\varphi{(J_z\cos{\nu}+J_y\sin{\nu})}-i\ham_{sqz}^{1a}t_{1a}\right]}.
\end{array}
\end{equation*}
Provided one can rotate the external field in the $z$-$y$ plane by an angle $\nu$, the squeezing can be performed while the field is on. The error caused by this procedure is $\propto \varphi Dt_{1a}\sin{(\nu)}\sim \varphi D/N$ for large $N$.
In the case of the two axis squeezing, as well, we can let the system rotate under the external field while squeezing provided the  field is rotated in the $x$-$y$ direction; the error we introduce in this case is $\propto \varphi D\log{(N)}/N$. 

 In Fig. (\ref{fig:etaSqz}) we compare the sensitivity achievable with an uncorrelated state to the sensitivity provided by a 1-axis and 2-axis squeezed state. In particular, we can identify a region of parameters where the squeezing sequence offer an advantage over the  simple echo  control. We assumed to have implanted with a varying density of NV centers a quasi-2D region of a diamond of volume $V=(300 \textrm{nm})^2$$\times$$10$nm (as this gives the best scaling with density).  NV densities between $4$$\times$$10^{14}$cm$^{-3}$ and $8$$\times$$10^{16}$cm$^{-3}$ occupying a small 2D layer could be obtained via geometrically controlled ion implantation~\cite{Schnitzler09} of high purity diamond. 
 
  Notice that although the requirements for squeezing seem to be daunting, since the time to obtain an optimal squeezing could be long, the control needed is not  more complex than what would be in any case required to  simply refocusing  the couplings. The  many-body protected manifold succeeds into protecting the squeezed state in the environment envisioned (mainly composed by nuclear spins), since the  noise correlation time is slow - on the order of millisecond as given by nuclear dipole-dipole interactions - while the couplings among NV centers can range up to hundreds of kHz because of the higher gyromagnetic ratio of electronic spins. In this situation,  the gap offers a good protection against the noise; with the current implantation techniques, however, the nuclear bath effects are overwhelmed by  the noise created by paramagnetic impurities, with much faster internal dynamics and we would expect  the protection to fail there.

\section{Conclusions}
We described how spin squeezed states can be created in dipolarly coupled electronic spin systems and used for precision measurement of external magnetic fields. The key features of the method proposed are its applicability even to spin systems with random couplings and an intrinsic protection against single-spin noise.
Although squeezed states are known to be more fragile to decoherence, in the present scheme the squeezing operator is in fact always accompanied by a many-body operator that provides protection to the squeezed state, reducing its dephasing rate to the rate of unsqueezed states. We studied the projected sensitivity gains for a particular application, an NV-based nanoscale magnetometer. As the control scheme needed to create an entangled state is not more demanding than the one required for the simple refocusing of the couplings, spin squeezing  will provide a practical sensitivity enhancement in very high quality materials.

\section{Acknowledgments}
We thank A.M. Rey and L. Jiang for stimulating discussions. This work was supported
by  ITAMP, the NSF, and the Packard Foundation.
\section{Appendix}
\subsection{Coherent averaging}
Multiple pulse sequences (MPS)  achieve the dynamical decoupling of unwanted interactions, or the creation of a  desired one, using coherent averaging. By
means of an external control the internal Hamiltonian is made
time-dependent; using cyclic MPS and
considering only stroboscopic measurements, the evolution is described
by an effective Hamiltonian that, to leading order in time, is given
by the time average of the modulated internal Hamiltonian. 
The evolution during a cycle can be better analyzed in the frame defined by the external control, where the internal Hamiltonian appears  time-dependent and periodic: $\tilde{\ham}_{int}(t)=U_c(t)^\dag\ham_{int}U_c(t)$. At times when $U_c(t)=\openone$, the evolution is given by $U(n_ct_c)=(\mathcal{T}e^{-i\int_0^{t_c} \tilde{\ham}_{int}(t) dt})^{n_c}$ (where $t_c$ is the cycle time and $n_c$ the number of cycles). The propagator can be rewritten using the Magnus expansion~\cite{Magnus}: 
\begin{equation}
\label{AHT}
	U(t_c)=\mathcal{T}e^{-i\int_0^{t_c} \tilde{\ham}_{int}(t) dt}=e^{-i[\bar{H}^{(1)}+\bar{H}^{(2)}+\dots..]t_c}, 
\end{equation}
where $\bar{H}^{(k)}$ are time-independent average Hamiltonians of order $k$~\cite{Haeberlen76}. The
MPS is tailored to produce the desired evolution usually up to the first or second order, and higher order terms lead to errors. 

In the case of the {\sc Mrev}8 sequence \cite{Mansfield71} shown in Fig. (\ref{fig:MREV8}) time symmetrization 
 brings to zero the second order terms \cite{Haeberlen76}. 
The leading order error $\bar{\ham}_{zz}^{(3)}$ cause dephasing of the spins. Its effects are captured by  a moment expansion \cite{Mehring} to second order of the effective Hamiltonian, $\langle
T_{\varphi,6}^2\rangle=\Tr{[\bar{\ham}^{(3)},J_\perp ]^2}/\Tr{J_\bot^2}$, where $J_\perp = \sum_k S_{\perp,k}$ is the collective spin in a direction perpendicular to $z$. 
$\langle T_{\varphi,6}^2\rangle$ has actually the character of a sixth moment and its value
is a function of the sixth power of the local field generated by the
dipolar interaction. The sensitivity decay rate is thus proportional to the sixth order of the density and the square of the total time, with a coefficient $\tilde{\alpha}=\langle T_{\varphi,6}^2\rangle/n_s^6\sim(\frac{\mu_0}{4\pi}\frac{(g\mu_B)^2}{\hbar})^6 \tau^4$. 
Note that the {\sc Mrev}8 sequence entails a large number of control
pulses.  For many typical errors (phase-lag and overshoot/undershoot)
the refocusing is only affected at higher order.  However,
depolarizing pulse errors occurring with probability $p$ lead to a
reduction of contrast: $C' = C (1-p)^k$ for $k$ pulses.  Using {\sc Mrev}8
with echo gives $k=34$ and a requirement $p \lesssim 0.002$ for
contrasts near unity.

\subsection{One axis-squeezing: analytical solution}
Here we provide an analytical solution to the one-axis squeezing dynamics~\cite{Kitagawa}, which has been used to calculate the behavior in the presence of dephasing noise.

The one axis squeezing operator reduces the variance of the collective magnetic moment along a direction at a variable angle $\nu$ in the y-z plane. Defining $\chi=d~t$, we obtain :
\begin{equation*}
	\langle J_x\rangle=\frac N2 \cos{\chi}^{N-1},\ \ \langle J_y\rangle=\langle J_z\rangle=0\\
\end{equation*}
\begin{equation*}
\Delta J_x^2=\frac N4[N-\frac{N-1}2P]-\left(\frac N2 \cos{\chi}^{N-1}\right)^2
\label{DeltaJZnu}
\end{equation*}
\begin{equation*}
	\Delta J_z^2(\nu)=\frac N4 \{1+\frac{N-1}4[P-\sqrt{P^2+Q^2}\cos{(2\nu+\atan{\frac QP})}]\}
\end{equation*}
where by $J_z(\nu)$ we indicate the operator $e^{i\nu J_x}J_ze^{-i\nu J_x}$ and we have set
\begin{equation*}
	\begin{array}{l}
	P=1-\cos^{N-2}{2\chi},\ \ \ 
	Q=4\sin{\chi}\cos^{N-2}{\chi}
\end{array}
\end{equation*}
The optimal value for $\nu$ (which minimize $\Delta J_z(\nu)$) is $\nu=-\half \atan{(Q/P)}$ [while $\nu=\pi/2-\half \atan{(Q/P)}$ would minimize $\Delta J_y(\nu)$]. Also notice that $\langle J_y(\nu)J_x\rangle=0$.

The squeezing parameter for this ideal case is:
\begin{equation}
	\xi^2=\frac{[1+\frac{N-1}4(P-\sqrt{P^2+Q^2})]}{(\cos{\chi}^{N-1})^2}
\end{equation}

For large spin systems and short times such that $N\chi^2\ll 1$ but $N\chi >1$, the optimal squeezing is obtained for $\chi_{1a}=\frac{3^{1/6}}{N^{2/3}}$, ($\nu\sim N^{-3}\approx0$)  and it scales with the number of spins as $\xi_{1a}=\frac{3^{1/3}}{\sqrt{2}N^{1/3}}$.

\bibliography{../../../Biblio}

\end{document}